\documentstyle[12pt]{article}
\begin{document}
\pagestyle{empty}
\centerline{\hfill UICHEP-TH/96-18}
\centerline{\hfill quant-ph/9611030}
\vskip 1in
\begin{center}
{\Large {\bf Accuracy of Semiclassical Methods}} \\
{\Large {\bf for Shape Invariant Potentials}}
\end{center}
\vspace{.2in}
\begin{center}
{\large {Marina Hr\v{u}ska\footnote{
e-mail: f2hruska@rcub.rcub.bg.ac.yu, \ 
Permanent address: Vlajkovi\'ceva 28, 11000 Belgrade, Yugoslavia.}, 
Wai--Yee Keung\footnote{e-mail: keung@uic.edu}
and Uday Sukhatme\footnote{e-mail: sukhatme@uic.edu}}}
\end{center}
\vspace{.1in}
\centerline {Department of Physics, University of Illinois at Chicago,}
\centerline {845 W. Taylor Street, Chicago, Illinois 60607-7059.}
\vspace{.8in}
\begin{abstract}
We study the accuracy of several alternative semiclassical methods 
by computing analytically 
the energy levels for many large classes of exactly 
solvable shape invariant potentials. For these potentials, the ground state 
energies computed via the 
WKB method typically deviate from the exact results by about 
$10\%$, a recently suggested modification using nonintegral Maslov 
indices is substantially better, and the supersymmetric WKB quantization 
method gives exact answers for all energy levels.
\end{abstract}
\newpage
\pagestyle{plain}
{\large {\bf 1. Introduction}}\vspace{.2in}

A variety of semiclassical methods have been proposed and used for 
determining the energy levels of one-dimensional potentials. The standard 
WKB method is discussed in most quantum mechanical 
texts \cite{Froman65,Landau65} but there are several recently suggested 
modifications based on supersymmetric quantum mechanics 
\cite{Comtet85}, energy dependent phase 
losses at the classical turning points with nonintegral Maslov indices 
\cite{Friedrich96}, and other related approaches \cite{Robbins90}.
As expected, all semiclassical
methods yield good energy eigenvalues 
$E_n$ for large values of the quantum number $n$. However, their accuracy 
varies quite substantially for small values of $n$ depending on the 
choice of potential. Also, many potentials only have a small number 
of bound states, so that the possibility of considering large 
values of $n$ does not exist. In this paper, we make a 
stringent test of the accuracy of various semiclassical methods by 
computing the eigenenergies of the ground state and other low lying 
states for many large classes of 
potentials which have the property 
of shape invariance \cite{Gendenshtein83}
under supersymmetry transformations. 
We have chosen shape invariant potentials since (i) they are exactly 
solvable and all eigenvalues are explicitly known; (ii) the integrals 
appearing in the semiclassical quantization conditions can all be 
performed analytically; (iii) the nonintegral Maslov indices used in the 
recent semiclassical approach \cite{Friedrich96} of Friedrich and Trost (FT) 
can be expressed in terms of superpotentials. Our plan is to first review 
three semiclassical approaches and the main ideas involving shape invariant 
potentials. We will then compute the quantization condition integrals 
analytically and determine energy eigenvalues. The results are 
presented in Tables 1 and 2, and we give some concluding remarks on the 
accuracy of various semiclassical approaches.\vspace{.2in}

{\large {\bf 2. Semiclassical Quantization Conditions}}\vspace{.2in}

{\bf (i) WKB Quantization:} 
The usual form of the semiclassical energy quantization condition (in 
units of $\hbar=2m=1$) is \cite{Froman65, Landau65}
\begin{equation} \label{eq1}
\int_{x_L}^{x_R} dx \sqrt{E-V(x)} = (n+\frac{\mu}{4})\pi~,
~\mu=\frac{\phi_L+\phi_R}{\pi/2}~,~~(n=0,1,2,...)~~,
\end{equation}
where the classical turning points $x_L$ and $x_R$ are given by
$V(x_L)=V(x_R)=E$.
The Maslov index $\mu$ denotes the total phase loss during one period 
in units of $\pi/2$. It contains contributions from the phase losses $\phi_L$
and  $\phi_R$ due to reflections at the left and right 
classical turning points $x_L$ and 
$x_R$ respectively. In the standard WKB approach, 
one takes $\phi_L=\phi_R=\pi/2$,
and an integer Maslov index $\mu=2$ for all energy levels. 
This gives the familiar result $(n+{1 \over 2})\pi$ for the right 
hand side of the WKB quantization condition.\vspace{.1in}

{\bf (ii) SWKB Quantization:} 
Another semiclassical approach \cite {Comtet85} which 
has been widely studied in recent years 
is based on the ideas of supersymmetric quantum mechanics\cite{Cooper95}. 
Here, the 
supersymmetric partner potentials $V_-(x)$ and $V_+(x)$ are 
given in terms of the superpotential $W(x)$ by $V_{\pm}=W^2(x) \pm W'(x)$. 
For the case of unbroken supersymmetry, $V_-(x)$ and 
$V_+(x)$ have degenerate energy levels 
except that $V_-(x)$ has an additional level at $E_0^{(-)}=0$. 
The corresponding ground state wave function $\psi_0^{(-)}(x)$ 
is related to the 
superpotential $W(x)$ via
\begin{equation} \label{eq2}
W(x)=-\frac{{\psi_0^{(-)}}'(x)}{\psi_0^{(-)}(x)}~~;
~~\psi_0^{(-)}(x) \propto~e^{-\int^x dx' W(x')} ~~~.
\end{equation}
The supersymmetric WKB (SWKB) approach \cite{Comtet85} results from 
combining the ideas of supersymmetry with the lowest order WKB method. The 
SWKB quantization condition is
\begin{equation} \label{eq3}
\int_{x'_L}^{x'_R} dx \sqrt{E^{(-)}-W^2(x)} = 
n \pi~,~(n=0,1,2,...)~~,
\end{equation}
where the two turning points ${x'_L}$ and ${x'_R}$ 
are given by $W(x)=\pm \sqrt{E^{(-)}}$. Note that for $n=0$, the turning points 
coincide and the SWKB quantization condition gives the exact result 
$E_0^{(-)}=0$ for the ground state energy.\vspace{.1in}

{\bf (iii) Friedrich-Trost Quantization:} 
This very recent 
proposal \cite{Friedrich96} makes use of the standard quantization 
condition [eq. (1)] with nonintegral, energy-dependent Maslov indices $\mu$. 
More specifically, the phase loss is taken to be given by 
\begin{equation} \label{eq4}
\tan(\frac{\phi_L}{2})=\frac{\psi'(x_L)}{k\psi(x_L)}~,~
\tan(\frac{\phi_R}{2})=-\frac{\psi'(x_R)}{k\psi(x_R)}~,
\end{equation}
where $k \equiv \sqrt{E-V_{min}}$ and $V_{min}$ is the minimum value 
of the potential $V(x)$.
In Ref. \cite{Friedrich96}, it was suggested that one 
could use the lowest order WKB 
wave function in eq. (\ref{eq4}) in order to determine the phase losses 
$\phi_L,\phi_R$ for any practical application. Indeed, it was shown that 
for power law potentials $x^p$ with $p=4,5,6$, this 
method gave better numerical results for the ground 
state energies than the standard WKB method, and also substantially 
improved wave functions. 

The FT approach of using the WKB wave function in eq. (\ref{eq4}) 
is rather cumbersome. It can be significantly simplified for the ground state 
by using eq. (\ref{eq2}) . The phase losses $\phi_L,\phi_R$ can then be 
re-written as
\begin{equation} \label{eq5}
\tan(\frac{\phi_L}{2})=-\frac{W(x_L)}{k}~,~
\tan(\frac{\phi_R}{2})=\frac{W(x_R)}{k}~.
\end{equation}
In this paper, we will assume eq. (5) to be valid for all energy levels 
in computing eigenenergies by the FT approach.\vspace{.2in}

{\large {\bf 3. Shape Invariant Potentials}}\vspace{.2in}

Given a superpotential $W(x,a_0)$ depending on a set of parameters $a_0$, 
the supersymmetric partner potentials $V_{\pm}(x,a_0)$  are given by 
\begin{equation} \label{eq6}
V_{\pm}(x,a_0)=W^2(x,a_0) \pm W'(x,a_0)~~.
\end{equation}
These partner potentials are shape invariant if they both 
have the same $x$-dependence Upton a change of parameters 
$a_1=f(a_0)$ and an additive constant $R(a_0)$. The shape invariance 
condition is 
\begin{equation} \label{eq7}
V_+(x,a_0)=V_-(x,a_1)+R(a_0)~~.
\end{equation}
This special property permits an immediate analytic determination of 
energy eigenvalues \cite{Gendenshtein83} and eigenfunctions \cite{Dutt86}. 
For unbroken supersymmetry, the eigenvalues are
\begin{equation} \label{eq8}
E_0^{(-)}=0~,~~E_n^{(-)}=\sum_{k=0}^{n-1} R(a_k)~.
\end{equation}
Many aspects of the bound states and scattering matrices of shape 
invariant potentials have been studied \cite{Cooper95} including 
several choices \cite{Khare93} for the change of parameters $a_1=f(a_0)$. 
In this paper, we confine our attention to shape invariant potentials 
corresponding to a translational change of parameters.\vspace{.2in}

{\large {\bf 4. Computation of Energy Eigenvalues}}\vspace{.2in}

All known families of shape invariant potentials in which the change of 
parameters is a translation $a_1=a_0+\beta$ are listed in Table 1. 
Names of the potentials and the corresponding superpotentials are given. 
Also tabulated is the minimum value $V_{-min}$ of the potential $V_-(x)$ 
and the position $x_{min}$ of the minimum. Subtracting $V_{-min}$ from 
$V_-(x)$  yields the tabulated potential $V(x)$, whose minimum value is 
clearly $V_{min}=0$. The exact energy eigenvalues $E_n^{exact}$ 
for $V(x)$ coming 
from eq. (\ref{eq8}) are also given.

To assess the accuracy of various semiclassical 
approaches, the first step is to evaluate the two types of integrals 
appearing in the quantization conditions. We denote the integral in the 
WKB condition eq. (1) by $I^{WKB}$ and the integral in the 
SWKB condition eq. (3) by $I^{SWKB}$. The integrals can be 
handled analytically using 
contour integration in the complex plane taking special 
care of the singularities at infinity and the cut going between the turning 
points. The results are given in Table 1 in terms of the following expressions:
$$
I_1(a,b) \equiv \int_{a}^{b} dy \sqrt{(y-a)(b-y)} = \frac{\pi}{8} (b-a)^2~~;
$$
$$
I_2(a,b) \equiv \int_{a}^{b} \frac{dy}{y} \sqrt{(y-a)(b-y)} = 
\frac{\pi}{2} (a+b)-\pi \sqrt{ab}~,(0<a<b)~;
$$
$$
I_3(a,b) \equiv \int_{a}^{b} \frac{dy}{y^2+1} \sqrt{(y-a)(b-y)} = 
\frac{\pi}{\sqrt{2}} [\sqrt{1+a^2}\sqrt{1+b^2}-ab+1]^{1/2}-\pi~~;
$$
$$
I_4(a,b) \equiv \int_{a}^{b} \frac{dy}{1-y^2} \sqrt{(y-a)(b-y)} = 
\frac{\pi}{2} [2-\sqrt{(1-a)(1-b)}-\sqrt{(1+a)(1+b)}]~,(-1<a<b<1)~;
$$
$$
I_5(a,b) \equiv \int_{a}^{b} \frac{dy}{y^2-1} \sqrt{(y-a)(b-y)} = 
\frac{\pi}{2} [\sqrt{(a+1)(b+1)}-\sqrt{(a-1)(b-1)}-2]~,(1<a<b)~;
$$
In all the above integrals, the limits $a,b$ are real numbers with 
$a<b$. We have given explicit expressions for the above integrals 
since they are not easily available in standard integration tables. Once 
$I^{WKB}$ and $I^{SWKB}$ have been computed, one can apply 
the quantization conditions to see how accurate the WKB, SWKB and FT approaches 
are. For the WKB and SWKB approaches, it is possible to get 
complete analytic results - these are shown in Table 1. The FT approach is also 
mostly analytical, but 
the final computations need numerical work. Results 
corresponding to specific numerical choices of the parameters 
appearing in the potentials are shown in Table 2. 

As an illustrative example, consider the Rosen-Morse II (hyperbolic) 
potential, for which the superpotential is 
$$
W(x)=A \tanh \alpha x + {B \over A}~.
$$
With a change of variables $y=\tanh \alpha x$, the SWKB integral is 
$$
I^{SWKB}= \frac{A}{\alpha} \int_{y'_L}^{y'_R} 
\frac {dy}{1-y^2} \sqrt {[-y^2-\frac{2B}{A^2}
+(\frac{E}{A^2}-\frac{B^2}{A^4})]}
$$
with turning points given by $A {y'}+{B \over A}= \pm \sqrt {E^{(-)}}$. 
One then sees that the integral is 
$\frac {A}{\alpha} I_4({y'_L},{y'_R})$. Substitution into 
the SWKB quantization 
condition $I^{SWKB}=n \pi$ and solving for $E^{(-)}$ gives
$$
E_n^{(-)SWKB}=A^2-{(A-n \alpha)}^2 +\frac{B^2}{A^2}-\frac{B^2}{{(A-n \alpha)}^2}
$$
which is the exact answer for all energy levels ! Similar steps give the 
WKB integral to be $I^{WKB}=\frac{\sqrt{A(A+\alpha)}}{\alpha}I_4(y_L,y_R)$ 
where the turning point are given by 
$$
A(A+\alpha)y^2+2By+(\frac{B^2}{A(A+\alpha)}-E)=0~~.
$$
Substitution into the WKB quantization condition $I^{WKB}=(n+{1 \over 2})\pi$ 
and solving for the energy gives
$$
E_n^{WKB} =A(A+\alpha)-{(\sqrt{A(A+\alpha)}-{\alpha \over 2}-n \alpha)}^2 
+\frac{B^2}{A(A+\alpha)}
-\frac{B^2}{{(\sqrt{A(A+\alpha)}-{\alpha \over 2}-n \alpha)}^2 }
$$
The WKB approach does not give the exact eigenvalues. The full energy 
computation for the FT quantization condition is harder to carry out 
analytically. 
For the numerical choice $\alpha =1, A=2, B=1$,
we see from Table 2 that the ground state 
energy $E_0^{FT}$ is lower than the exact energy by $3.2\%$ whereas 
$E_0^{WKB}$ is higher than the exact energy by $9.7\%$.\vspace{.2in}

{\large {\bf 5. Conclusion}}\vspace{.2in}

We have given a complete analytic treatment of the energy levels 
of shape invariant potentials (with a translational change of parameters) 
for various semiclassical quantization conditions. As expected from 
previous work \cite{Dutt86}, the SWKB energy levels are exact. The WKB 
energy levels are exact for the one-dimensional harmonic oscillator and the 
Morse potentials only. For other potentials the results are not exact, and 
this has historically led to ad hoc Langer corrections \cite{Langer37}. 
Typically, one sees 
from Table 1 that the 
ground state energy from the WKB method deviates from the exact result by 
about $10\%$. The new semiclassical approach of Friedrich and Trost is 
exact only for the one-dimensional harmonic oscillator and no other shape 
invariant potential. However, in general, it gives significantly more accurate 
ground state energies than the WKB method.\vspace{.2in}

This work was supported in part by the U. S. Department of Energy.
\vspace{1in}

\noindent
{\large {\bf Table Captions}}\vspace{.2in}

\noindent
{\bf Table 1:} List of all shape invariant potentials 
and their eigenvalues. Analytic 
expressions for the integrals in the WKB and SWKB quantization condition are 
given, along with the energy eigenvalues. For these potentials, the SWKB 
results are always exact, whereas the WKB results are exact only for the 
harmonic oscillator and Morse potentials.

\noindent
{\bf Table 2:} Comparison of the exact ground state energies of shape invariant 
potentials with results from the WKB and Friedrich-Trost method. The percent 
errors are also shown. The SWKB results are not shown, since they are always 
exact for the potentials under consideration.

\end{document}